\documentclass[12pt]{iopart}

\newcommand{\be}{\begin{equation}}
\newcommand{\ee}{  \end{equation}}
\newcommand{\ba}{\begin{eqnarray}}
\newcommand{\ea}{  \end{eqnarray}}

\begin{document}

\title{Open Problems in Applying Random-Matrix Theory to Nuclear Reactions}

\author{H. A. Weidenm\"{u}ller}

\address{Max-Planck-Institut f{\"u}r Kernphysik, 69029 Heidelberg, Germany}
\ead{Hans.Weidenmueller@mpi-hd.mpg.de}
\begin{abstract}

Problems in applying random-matrix theory (RMT) to nuclear reactions
arise in two domains. To justify the approach, statistical properties
of isolated resonances observed experimentally must agree with RMT
predictions. That agreement is less striking than would be desirable.
In the implementation of the approach, the range of theoretically
predicted observables is too narrow.

\end{abstract}



\submitto{\JPG}

\maketitle

\section{Purpose}
\label{purp}

Nuclear Reactions on medium-weight and heavy target nuclei in the
low-energy domain (bombarding energies up to several tens of MeV) are
dominated by compound-nucleus (CN) resonances. An adequate reaction
theory requires information on the spins, parities, partial and total
widths and on the spacings of these resonances. With typical resonance
spacings $d$ (for fixed quantum numbers) in the $10$ eV range even at
neutron threshold (and with very much smaller values of $d$ at higher
excitation energies), nuclear-structure theory cannot supply such
information. The neccessary calculations are beyond present-day
capabilities. In its stead one follows Wigner's~\cite{Wig55, Wig57}
original proposal and uses random-matrix theory (RMT). RMT in its
time-reversal invariant form (the Gaussian Orthogonal Ensemble (GOE))
has become a standard tool of nuclear reaction theory~\cite{Wei09,
  Mit10}.

We ask: What are the problems and challenges in applying the GOE to
nuclear reactions? The question has two parts. (i) Are the properties
predicted by the GOE consistent with experimental data on CN
resonances? (ii) Is it possible to implement the GOE into a viable and
useful theory of nuclear resonance reactions? An affirmative answer to
question (i) is the neccessary condition for a physically meaningful
use of RMT in nuclear reaction theory. Asking that question is
particularly timely. Refs.~\cite{Koe10, Koe11, Koe12} reported strong
disagreement of neutron resonance data with GOE predictions. Some of
these results have found wide attention~\cite{Rei10, Wei10, Cel11,
  Vol11} eroding, as they seemingly do, one of the cornerstones of the
statistical theory of nuclear reactions~\cite{Wei09, Mit10}. After
defining the GOE in Section~\ref{goe} we address questions (i) and
(ii) in turn.

\section{The GOE}
\label{goe}

The GOE is an ensemble of Hamiltonian matrices $H$ in $N$-dimensional
Hilbert space. Because of time-reversal invariance, the matrices $H$
are real and symmetric. The ensemble is defined in terms of the
probability density
\be
{\cal P}(H) {\rm d} H = {\cal N}_0 \exp \bigg\{ - \frac{N}{4 \lambda^2}
{\rm Tr} (H^2) \bigg\} \ {\rm d} H \ . 
\label{1}
\ee
Here ${\cal N}_0$ is a normalization constant, $\lambda$ is a
parameter of dimension energy which (in the center of the GOE
spectrum) is related to the mean level spacing $d$ by $d = \pi \lambda
/ N$, and
\be
{\rm d} H = \prod_{\mu \leq \nu}^N {\rm d} H_{\mu \nu}
\label{2}
\ee
is the product of the differentials of the independent matrix
elements. Eq.~(\ref{1}) shows that the independent elements of $H$ are
uncorrelated Gaussian-distributed random variables. Eqs.~(\ref{1}) and
(\ref{2}) show that the ensemble is invariant under orthogonal
transformations of Hilbert space. Hence its name Gaussian Orthogonal
Ensemble.

Every matrix $H$ can be diagonalized by an orthogonal transformation
${\cal O}$. The eigenvalues are denoted by $E_\mu$ with $\mu = 1,
\ldots, N$. In terms of these variables the distribution takes the
form
\ba
{\cal P}(H) {\rm d} H = {\cal N}_0 \ {\rm d} {\cal O} \ \exp \bigg\{
- \frac{N}{4 \lambda^2} \sum_\mu E^2_\mu \bigg\} \prod_{\rho < \sigma}
|E_\rho - E_\sigma| \prod_\nu {d} E_\nu \ .
\label{3}
\ea
The factor ${\rm d} {\cal O}$ is the Haar measure of the orthogonal
group in $N$ dimensions and defines the distribution of eigenvectors.
It implies that in the limit $N \to \infty$ the projections of all
eigenvectors onto any fixed vector in Hilbert space are
Gaussian-distributed real random variables. The factor $\prod_{\rho <
  \sigma} |E_\rho - E_\sigma|$ causes the linear level repulsion
characteristic of the GOE. The distribution~(\ref{3}) factorizes, one
factor depending only on the eigenvectors and the other, only on the
eigenvalues. Hence, eigenvectors and eigenvalues are statistically
independent. This result follows directly from the orthogonal
invariance of the GOE.

Theoretical predictions of the GOE for spectral fluctuations
(Section~\ref{spec}) and for cross-section fluctuations
(Section~\ref{reac}) are based on ensemble averages. These are
compared with experimental data obtained by taking the running average
over energy of a single spectrum. For the GOE it has been shown
(albeit only for a restricted set of observables) that the ensemble
average and the running average taken over the spectrum of a single
member of the GOE agree. This fact is the basis for comparison with
experimental data. To minimize the statistical error, long data
sequences are required.

According to the Bohigas-Giannoni-Schmit conjecture~\cite{Boh84}, the
spectral fluctuation properties of dynamical systems that are chaotic
in the classical limit agree with RMT predictions. Therefore, the
applicability of the GOE to nuclear reaction theory is linked to the
question whether the nuclear dynamics is chaotic. We do not discuss
here general tests relevant to that question (such as the spacing
distribution of levels in the ground-state domain, or of levels some
$100$ keV above the yrast line, or of the eigenvalues of the
interacting boson model). The dynamics of such low-lying excitations
do not have any immediate bearing on nuclear reaction theory.

\section{Tests of GOE Spectral Fluctuations in Nuclei}
\label{spec}

Three basic predictions of the GOE can be tested: (i) the distribution
of level spacings as obtained from the last three factors in
Eq.~(\ref{3}) in the limit $N \to \infty$; (ii) the Gaussian
distribution of eigenvectors deduced in the same limit, (iii) the
independence of the distributions of eigenvectors and eigenvalues.
Significant tests have been done either on data generated by
large-scale shell-model calculations, or on experimental data obtained
from the analysis of sequences of isolated resonances. To be
statistically significant, the data sets should be large, each set
should be clean (the quantum numbers of all resonances should be
unambiguously known and be identical), and the sets should be complete
(no resonances missing). The three requirements are difficult to
fulfill experimentally (long sequences are hard to come by, very
narrow resonances are easily missed, unambiguous spin assignments may
be very difficult to obtain). In spite of $60$ years of work on the
problem, the experimental data sets that fulfill these requirements
are, therefore, small in number. Results of shell-model calculations
have been extensively tested for GOE properties only for $s-d$-shell
nuclei.

\subsection{Level-Spacing Distribution}

Two measures are mainly used to test GOE predictions for the
level-spacing distribution: Wigner's surmise for the nearest-neighbour
spacing (NNS) distribution, and the $\Delta_3$ statistic of Mehta and
Dyson. The NNS distribution gives the distribution of spacings of
neighbouring eigenvalues. It accounts mainly for level repulsion and
not for long-range correlations of spacings. It is not a very
sensitive test of general GOE properties. Numerical studies of
few-degrees-of-freedom systems with varying interaction strength have
shown that level repulsion sets in quite early and before all other
GOE properties are attained. The $\Delta_3$ statistic is more
sensitive as it measures long-range correlations of level spacings. It
requires long sequences of data.

Results of shell-model calculations for $s-d$-shell nuclei were tested
for agreement with GOE predictions in Ref.~\cite{Zel96}. Both the NNS
distribution and the $\Delta_3$ statistic were calculated for the 3276
states with spin $J = 2$ and isospin $T = 0$ that occur in the middle
of the shell. In both cases, agreement with the GOE is very good,
except that the $\Delta_3$ statistic has a tendency to lie slightly
above the GOE curve for very large lengths $L$. That tendency seems
not fully understood. The agreement is expected. It is due to the
(almost) complete mixing of shell-model configurations in eigenstates
of the shell-model Hamiltonian that are near the center of the
spectrum. To be sure, the mixing addressed in Ref.~\cite{Zel96} refers
only to the configurations within a single major shell. Shell
structure being manifest over a wide range of excitation energies, one
may wonder what happens at excitation energies that lie in the middle
between two major shells. That question seems not to have been
addressed in the theoretical literature. With this proviso, it is
theoretically expected that CN resonances follow the GOE.

Relevant experimental information comes from the analysis of isolated
resonances measured in the scattering of slow neutrons on
medium-weight and heavy nuclei and, with a smaller data set, from
scattering of protons on medium-weight nuclei at energies near the
Coulomb barrier. For neutrons scattered on even-even nuclei, all
$s$-wave resonances have spin/parity assignment $1/2^+$ while $p$-wave
resonances carry spin-parity assignments $1/2^-$ and $3/2^-$. Because
of the angular-momentum barrier, the latter are suppressed by a factor
$E$ (the resonance energy measured from threshold) relative to the
former. Spin-parity assignments for resonances in proton scattering
are frequently less problematic. Most available sequences of
resonances being quite short, Haq, Pandey and Bohigas~\cite{Haq82} in
the 1980's combined resonance data from a number of nuclei into the
``Nuclear Data Ensemble'' (NDE). The NDE comprises $30$ sequences of
levels of $27$ different nuclei amounting to a total of $1407$
resonance energies~\cite{Haq82}. Up to this day, the NDE or a suitable
modification that takes into account more recent data has been the
backbone of work on the level-spacing distribution.

Under the (tacit) assumption that all neutron resonances in the NDE
have spin/parity assignment $1/2^+$ (i.e., no $p$-wave admixtures), it
was shown in Ref.~\cite{Haq82} that the data were in good agreement
with the $\Delta_3$ statistic for sequence lengths $L \leq 20$. In
Ref.~\cite{Boh83} the analysis was extended to the NNS distribution
(resulting in good agreement with the Wigner surmise), and to the
Porter-Thomas distribution (discussed below) for neutron widths.  The
$\Delta_3$ statistic is determined by the two-level correlation
function. Measures accounting for the three- and four-level
correlation functions were evaluated for the NDE and compared with GOE
predictions in Ref.~\cite{Boh85}, again resulting in good agreement.
A further test~\cite{Lom94} involving higher $n$-point functions uses
the fact that omission of every other level in a GOE spectrum yields a
GSE spectrum (the GSE is the Gaussian symplectic ensemble of random
matrices). In applying that test to the NDE, ``no disagreement with
the GOE'' could be detected~\cite{Lom94}. Taken together, the work of
Refs.~\cite{Haq82, Boh83, Boh85, Lom94} yielded very strong evidence
for the agreement of nuclear spectral fluctuations with GOE
predictions.

The only existing test for correlations between eigenvalues and
eigenfunctions seems to be the one in Ref.~\cite{Lom94}. The NDE gave
no evidence for such correlations, in agreement with the GOE.

One cause of uncertainty in the data analysis described so far is due
to the possible presence of $p$-wave resonances in low-energy neutron
scattering data. It is clear that firm and unbiased conclusions can
only be drawn if $s$-wave and $p$-wave neutron resonances are cleanly
separated. The problem was emphasized by Koehler who reconstructed and
reanalysed the NDE~\cite{Koe11}. Reconstruction was necessary because
a full account of the analysis in Refs.~\cite{Haq82, Boh83} of the NDE
was never published, and because some of the references therein were
based on private communication. The reconstructed NDE contains $1245$
resonances and is not completely identical with the original one. A
significant fraction (typically $5 - 10$ percent) of the neutron
resonances are of unknown parity or known to be $p$ wave.

The $\Delta_3$ statistic for the reconstructed NDE was analysed in
Ref.~\cite{Koe12}. Upon first sight, agreement with the GOE is
obtained. The $\Delta_3$ statistic using all nuclei in the
reconstructed NDE agrees with the GOE, confirming the conclusion of
Ref.~\cite{Haq82}. Moreover, the $\Delta_3$ statistic for the two
nuclides with the longest level sequences (U and Th nuclei) also
agrees with the GOE. However, for the remaining nuclei in the
reconstructed NDE (U and Th nuclei omitted) the $\Delta_3$ statistic
deviates increasingly with increasing length $L$ from the GOE
prediction, the deviation becoming significant for lengths $L > 20$ or
so. A similar deviation is found when resonances with a definite
$p$-wave assigment are removed. These are mainly resonances in U and
Th. It is concluded that contrary to the first impression, the
$\Delta_3$ statistic for the reconstructed NDE is in serious
disagreement with the GOE.

Taken by themselves, the results on the $\Delta_3$ statistic in
Ref.~\cite{Koe12} suggest rejection of the GOE hypothesis. In its
entirety, however, work on the level-spacing distribution suggests a
different conclusion. The test of the $\Delta_3$ statistic in
Ref.~\cite{Haq82} was confined to sequence lengths $L \leq 20$. In
that domain both Ref.~\cite{Haq82} and Ref.~\cite{Koe12} show full
agreement with GOE predictions. The higher-order correlations tested
in Refs.~\cite{Boh85, Lom94} also agree with the GOE. It is very
difficult to see how such agreement could survive the admixture of
$p$-wave resonances, or be accidental. For instance, it appears
extremely unlikely that a sequence of $s$-wave resonances with some
accidental admixture of $p$-wave resonances would, upon omission of
every second resonance, obey GSE statistics. Given these facts we are
left with the disagreement with GOE predictions found in
Ref.~\cite{Koe12} for the $\Delta_3$ statistic and for sequence
lengths $L > 20$. Here several questions come to mind. (i) Table I of
Ref.~\cite{Koe11} shows that the number of sequences with $L > 20$
decreases strongly with increasing $L$. Estimates of the statistical
uncertainty of the $\Delta_3$ statistic in Ref.~\cite{Koe12} do not
seem to reflect that fact. (ii) It is known that with increasing
distance from neutron threshold, an increasing number of narrow
resonances is missed experimentally. The remaining resonances lack the
stiffness of the spectrum and are, therefore, expected to be more
random. The resulting deviations of the $\Delta_3$ statistic from the
GOE prediction should tend towards the Poisson distribution. This is
what the data show.  (iii) Before removal of the supposed $p$-wave
resonances, the $\Delta_3$ statistic for U and Th agrees with the
$\Delta_3$ statistic for the GOE. After removal of these resonances,
the $\Delta_3$ statistic for the remaining $s$-wave resonances does
not. It would take a very subtle correlation between $s$-wave and
$p$-wave resonances to re-establish GOE properties in the combined
sequence when the $s$-wave resonances alone lack such properties. The
accidental occurrence of such a correlation seems utterly improbable.
A dynamical correlation is excluded because states with different
quantum numbers do not interact and, thus, lack the mechanism that
would cause stiffness of the combined spectrum. An erroneus $p$-wave
assignment to the excluded resonances seems to offer the most likely
explanation.

\subsection{Width Distribution}
\label{wid}

The partial-width amplitude of a CN resonance is proportional to the
overlap of the resonance wave function with the channel wave function.
In the RMT approach the resonance wave function is an eigenfunction of
the GOE. Then, the partial-width amplitude is the projection of a GOE
eigenvector unto some fixed vector in Hilbert space. Therefore, the
GOE predicts that the partial-width amplitudes of CN resonances have a
Gaussian distribution. Such amplitudes can only be measured in special
cases. Relevant information comes mainly from neutron widths of
isolated CN resonances measured in slow neutron scattering. The
neutron width being the square of the neutron partial-width amplitude,
its distribution is predicted to be the Porter-Thomas distribution
(PTD) (a $\chi^2$ distribution with a single degree of freedom, $\nu =
1$).

The shell-model calculations~\cite{Zel96} referred to above for states
with $J = 2$ and $T = 0$ in the middle of the $s-d$-shell have also
been tested for comparison with GOE predictions for the
eigenfunctions. Good agreement was found. That is another
demonstration of the (nearly) complete mixing of shell-model
configurations caused by the residual interaction. As in the case of
the spectral fluctuations, the theoretical expectations are that GOE
predictions apply to CN resonance eigenfunctions.

Experimental data on partial-width amplitudes come from proton
scattering on medium-weight nuclei at energies near the Coulomb
barrier, those on partial widths from data on isolated CN resonances
in slow neutron scattering.

A very thorough test of the postulated Gaussian distribution of
partial-width amplitudes was reported in Ref.~\cite{Shr87}. Data on
proton scattering by medium-weight nuclei yielded $1117$ reduced
partial-width amplitudes. If the amplitudes have a Gaussian
distribution, the linear correlation coefficient of the squares of the
amplitudes must be equal to the square of the correlation coefficient
for the amplitudes. The positive test~\cite{Shr87} provides strong
support for the Gaussian distribution.

In the original analysis of the NDE, the distribution of neutron
widths showed good agreement with the PTD~\cite{Boh83}. As in the case
of the level distribution, it is important to ascertain that all
levels included in the analysis are $s$-wave resonances. A method to
exclude $p$-wave resonances (as well as to avoid $s$-wave resonances
with very small widths that partly escape experimental detection
anyhow) in neutron resonance data was introduced in Ref.~\cite{Cam94}.
With $E$ the resonance energy counted from threshold, $s$-wave
($p$-wave) resonances have an intrinsic energy dependence $E^{1/2}$
($E^{3/2}$, respectively). The transition to reduced widths removes
the $E^{1/2}$ dependence of the $s$-wave resonances and leaves a
linear $E$ dependence of the $p$-wave resonances. The latter (and all
very narrow $s$-wave resonances) were removed by a cutoff. Only
resonances with widths larger than the cutoff were retained. For a set
of $9$ nuclides the cutoff was chosen differently for each nucleus and
in each case amounted to less than $10$ percent of the average
width. A maximum-likelihood (ML) analysis was used to test whether the
remaining data were in agreement with the PTD. The likelihood function
was dependent on $\nu$, the number of degrees of freedom of a
$\chi$-squared distribution, and on the average width $\langle \Gamma
\rangle$. For each nucleus, both $\nu$ and $\langle \Gamma \rangle$
were determined by the maximum of the likelihood function. The
resulting values of $\nu$ were found to depend on the cutoff and, for
the set of $9$ nuclides, ranged from $0.64 \pm 0.28$ to $1.32 \pm
0.30$, with an error-weighted mean value $\nu = 0.98 \pm 0.10$. In
Ref.~\cite{Cam94} that result was considered to be ``completely
consistent with the PTD''.

In the ML analysis of the width distribution for the reconstructed NDE
in Ref.~\cite{Koe11}, a modified version of the cutoff procedure
introduced in Ref.~\cite{Cam94} was employed. A linearly
energy-dependent cutoff was used. That cutoff safely removes all
$p$-wave resonances. In addition, it simulates the experimental
tendency to miss with increasing energy an increasing number of narrow
$s$-wave resonances. The ML analysis~\cite{Cam94} gave $\nu$ values
for the individual nuclides (denoted by $\nu_{pf}$ in Table I of
Ref.~\cite{Koe11}) that range widely from $0.49 (+ 0.64, - 0.48)$ to
$3.6 (+ 1.6, - 1.3)$. The resulting weighted average is $\nu = 1.217
\pm 0.092$. This result rejects the PTD with a statistical
significance of at least $98.17$ percent.

The same approach was used in Ref.~\cite{Koe10} to analyse a set of
neutron resonance data in the Pt nuclei ($158$ resonances for
$^{192}$Pt, $411$ resonances for $^{194}$Pt). The ML analysis with a
cutoff yields very small $\nu$ values ($\nu = 0.47$ for $^{192}$Pt,
$\nu = 0.60$ for $^{194}$Pt). When combined, these results reject
agreement with RMT with a statistical significance of at least
$99.997$ per cent probability~\cite{Koe10}. These results caused some
theoretical activity. However, none of the suggestions put forward in
Refs.~\cite{Wei10, Cel11, Vol11} seems able to remove that
discrepancy~\cite{Koe11a} with the GOE.

The conclusions in Refs.~\cite{Koe10, Koe11} are based on the ML
analysis with a cutoff. How reliable is that method? In
Ref.~\cite{Shr14} the question is addressed both analytically and with
the help of computer simulations. Starting point is the PTD with unit
width. A fictitious ensemble of neutron widths is generated by drawing
$N$ widths randomly from the PTD, and by repeating the procedure
$2500$ times. By construction, these widths follow the PTD. The data
are analysed with the help of an ML function with a cutoff. The
procedure does not generate resonance energies so the cutoff chosen is
constant. For each member of the ensemble, the maximum of the ML
function yields a value for $\nu$. The distribution of the resulting
$\nu$ values over the ensemble is investigated numerically. For a
cutoff of $0.1$ ($10$ percent of the average width) and $M = 100$, the
histogram for the distribution resembles a Gaussian centered at $\nu
\approx 1$. The full width at half maximum is approximately unity. The
combined error due to cutoff and finite $M$ value is, thus, much
bigger than the typical error estimated from the width of the maximum
of the ML function for a single member of the ensemble. In
Ref.~\cite{Shr14} the origin of that error and its substantial growth
with increasing cutoff are displayed. As a result, the ML analysis of
a single set of widths drawn from a PTD may yield a value for $\nu$
that differs widely (by $\pm 1/2$ or so) from the actual value $\nu =
1$.

In the light of these results, it is not clear whether the two small
$\nu$ values found in Ref.~\cite{Koe10} are inconsistent with the
GOE. Moreover, it is conceivable that the spread of $\nu$ values found
for a set of nuclides in Ref.~\cite{Cam94}, and for the reconstructed
NDE in Ref.~\cite{Koe11}, both reflect the width of the distribution
of $\nu$ values due to cutoff and finite sample size. It appears that
definite conclusions on the validity of the GOE hypothesis can be
drawn only when every ML analysis using a cutoff is supported by
simulations of the type used in Ref.~\cite{Shr14}.

In summary, there is substantial theoretical~\cite{Zel96} and
experimental~\cite{Shr87} evidence in favour of a Gaussian
distribution of partial-width amplitudes. The case of neutron
widths~\cite{Koe10, Koe11} is unresolved.

\section{Implementation of the GOE for Nuclear Reactions}
\label{reac}

We turn to question (ii) raised in Section~\ref{purp}: How is it
possible to implement the GOE into a viable and useful theory of
nuclear resonance reactions? In the standard approach and in the
absence of direct reactions, the symmetric and unitary scattering
matrix is written as~\cite{Wei09, Mit10}
\be
S_{a b}(E) = \delta_{a b} - 2 i \pi (W^\dag D^{- 1}(E) W)_{a b}
\label{4}
\ee
where
\be
D_{\mu \nu}(E) = E \delta_{\mu \nu} - H_{\mu \nu} + i \pi (W W^\dag)_{\mu
\nu} \ .
\label{5}
\ee
Here $E$ is the energy. The indices $a, b$ ($\mu, \nu$) denote the
$\Lambda$ open channels (the $N$ states of the GOE, respectively).
The Hamiltonian $H$ is a member of the GOE defined in
Section~\ref{goe}. The matrix $W$ is rectangular. The elements $W_{\mu
  a}$ couple the GOE states to the channels. Because of time-reversal
invariance, the $W_{\mu a}$ are real so that $W_{\mu a} = W_{a \mu}$,
they are independent of energy $E$, and they obey $(W^\dag W)_{a b} =
\delta_{a b} N v^2_a$. The last condition guarantees diagonality of
the average $S$ matrix, i.e., absence of direct reactions. The form of
$S$ in Eqs.~(\ref{4}, \ref{5}) seems uncontroversial and generally
accepted. The same is true for a more general form of $S$ that allows
for the presence of direct reactions~\cite{Wei09, Mit10}. Equivalent
forms (where $S$ is written in terms of the $K$ matrix) are also
commonly used. In the limit $N \to \infty$ the number of parameters of
the GOE approach is $\Lambda$: the spacing parameter $\lambda$ and the
$\Lambda$ parameters $v^2_a$ for the coupling strengths to the
channels are combined to yield $\Lambda$ dimensionless parameters
$v^2_a / \lambda$. That number is equal to the number of diagonal
average $S$-matrix elements. The latter serve as input for the
prediction of $S$-matrix fluctuation properties.

The number $\Lambda$ of open channels in Eqs.~(\ref{4}, \ref{5}) is
considered fixed. With increasing energy, the number of open channels
in any nuclear reaction actually increases exponentially. Therefore,
Eqs.~(\ref{4}, \ref{5}) provide a useful approximation to $S$ only
within some finite energy interval $\Delta E$. The interval is defined
by the condition that the coupling of GOE states to all channels with
thresholds within $\Delta E$ is sufficiently weak. Under that
condition the neglect of the energy dependence of the matrix $W$ is
also justified. The assumption is most strongly violated by $s$-wave
neutron channels because here neither angular momentum barrier nor
Coulomb barrier suppress the coupling to the GOE states. It seems that
the number of such cases that are experimentally relevant, is small.
Their adequate theoretical treatment would pose a problem.

In principle, Eqs.~(\ref{4}) and (\ref{5}) can be used to predict
theoretically correlation functions (as ensemble averages) involving
products of two or more $S$-matrix elements. Because of the technical
difficulties in the calculation of such averages, the predictive power
of the GOE for nuclear reactions is actually much more limited than
for spectral fluctuations. General analytical results exist only for
the correlation function involving a pair of $S$-matrix
elements~\cite{Ver85}, for select values of the correlation function
involving three or four $S$-matrix elements~\cite{Dav88, Dav89}, and
for the probability distribution of single $S$-matrix
elements~\cite{Fyo05, Roz03, Roz04, Kum13}. The complete joint
probability distribution of all $S$-matrix elements is
known~\cite{Plu13a, Plu13b, Plu13c} only in the Ericson regime
(strongly overlapping resonances). From a practical point of view,
results beyond average cross sections and beyond the Ericson regime,
especially for cross-section correlation functions, would be of
considerable interest. The theoretical treatment of direct reactions
(non-zero values of non-diagonal average $S$-matrix elements) does not
seem to pose a problem.

\section{Summary}

The problems in applying random-matrix theory to nuclear reactions lie
in two domains. (i) In the justification of the approach. Such
justification must be based upon established statistical properties of
isolated resonances. The data base is narrow. In general, the data
agree with GOE predictions. Open questions exist for the $\Delta_3$
statistic for long sequences of levels, and for the distribution of
neutron widths. (ii) In the implementation of the approach. The range
of theoretically predicted observables has grown in recent years but
is still too narrow. Theoretical expressions for cross-section
correlation functions would be particularly valuable.

\section*{Acknowledgments}
The author is grateful to J. F. Shriner Jr. and to G. E. Mitchell for
many useful communications relating to aspects of nuclear reaction
theory.

\end{document}